\documentclass[aps,prb,twocolumn,showpacs]{revtex4}
\usepackage{amsmath,amsthm}
\usepackage{epsf}
\usepackage{amssymb}
\usepackage{graphicx}
\usepackage{color}      
\usepackage{psfrag}

\begin{document}
\title{Steering of a Bosonic Mode with a Double Quantum Dot}
\author{T. Brandes, N. Lambert}
\affiliation{Department of Physics, University of Manchester
Institute of Science and Technology (UMIST), P.O Box 88,
Manchester M60 1QD, United Kingdom}
\date{\today}
\begin{abstract}
We investigate the transport and coherence
properties of a double quantum dot coupled to a single damped boson mode.
Our numerically results reveal how the
properties of the boson distribution can be steered by altering parameters 
of the electronic system such as the energy difference between the dots.
Quadrature amplitude variances and the Wigner function are employed to illustrate
how the state of the boson mode can be controlled by a stationary electron current through the dots.
\end{abstract}
\maketitle

\section{Introduction}
The two-level system coupled to a single bosonic mode (Rabi Hamiltonian \cite{Rabi37}) is probably one of the
best studied models for the interaction of matter with light \cite{Allen}.
Cavity quantum electrodynamics is an example where the coupling between
atoms and photons can be studied in detail and used in order to, e.g., 
transfer quantum coherence from light to matter
(control of tunneling by electromagnetic fields \cite{NS96,GH98}) and vice versa \cite{UWK96,SU97,SU99}.

Quantum optics usually deals with {\em closed} two-level systems where the
total electron number on individual atoms remains constant
and does not fluctuate during the interaction with the photon.
This restriction can be lifted 
in semiconductor quantum dots (artificial atoms) 
by tunnel coupling to electron reservoirs. For example,
in semiconductor cavity quantum electrodynamics \cite{Yamamoto} the interaction of light with excitons
can be steered by `pumping' the cavity by resonant tunneling of electrons and holes.

In this paper, we examine a single bosonic mode interacting with a
system of two (bound) electronic states which are themselves coupled to a continuum of free electrons. As a
concrete realization we investigate the stationary electron transport through
a double quantum dot coupled to electron reservoirs and a single
photon or phonon mode. The main idea is to control the density
matrix of the coupled dot-boson system by external parameters such
as the reservoir chemical potentials and the tunnel couplings.
These parameters (which in experiments can be controlled via gate
voltages) then determine the reduced density operators of the dot
and the boson and allow one, e.g.,  to modify the state of the boson 
by driving a stationary electron current through the dots.

One of the motivations for our work are experiments in double quantum dots where the
coherent coupling between classical light (microwaves) and electrons
can be detected in electronic transport.
\cite{SN96,Oosetal98,Blietal98b,Blietal98a}. The open two-level-boson model
describes the full quantum version of these systems with
the light replaced by a photon field with its own quantum dynamics.

On the other hand, quantum effects (like spontaneous emission)
relevant for transport through double dots have been found to be due to {\em phonons} rather than photons.
In fact, 
the importance of electron-phonon coupling for 
transport in coupled dots is well-established by now
\cite{Fujetal98,Taretal99,BK99,Qinetal01,Fujetal02}. Suggestions have been
made \cite{Fujetal98,WI02,DBK02} to explore
`semiconductor phonon cavity QED'  in
nanostructures where phonons become experimentally controllable. 
In contrast to standard
semiconductor cavity  QED \cite{Yamamoto}, boundary conditions for {\em vibration} modes
lead to extremely non-linear phonon dispersion
relations. For example, van-Hove singularities in the phonon DOS show up at certain 
frequencies \cite{DBK02}, signatures of which
seem to be relevant for transport through quantum dots embedded into free-standing
structures \cite{Hoeetal02}. Such a situation would then  be described (within a  strongly idealized model)
by the coupling of a single frequency boson mode to a few-level quantum dot.

The influence of `nano-mechanical' vibrational properties on single electron tunneling in fact has emerged
as a whole new area of mesoscopic transport, triggered by the possibility to 
explore electron transport through individual molecules \cite{Paretal00,JGA00,Reietal02,BS01,GK02} or in 
freestanding \cite{SCCR95,CR98,BRWB98,Blietal00} and movable \cite{Goretal98,WZ99,AWZB01,AM02} nano-structures.
Phonons then are no longer a mere source of dissipation but become experimentally controllable with
the possibility, e.g.,  to create phonon confinement \cite{BMS94,BAMS95,NAW97,DBK02} 
or the analogon of quantum optical
phenomena such as coherent and squeezed phonon states. \cite{HN97,KW98,Blen99,BW00}

The paper is organized as follows:
in section \ref{section_model}, we introduce the model and derive a master equation for the
density operator. We discuss transport properties in
section \ref{section_transport} where a comparison to analytical solutions is made.
In section \ref{section_boson}, the reduced boson density operator, its Wigner function and
the fluctuation properties of the boson mode under a stationary electron current are discussed.
Finally, we conclude with a discussion in section~\ref{section_discussion}.

\section{Model and Method}\label{section_model}
We consider an idealized situation of a single cavity boson mode (photon
or phonon) coupled to a  two-level electronic system (dot) 
which itself is connected to external electron reservoirs. 
Our aim is to determine the stationary reduced density operator $\rho(t)$ of the 
dot-boson system for large times $t$, treating the dot and the boson 
on equal footing. 
The single bosonic mode (photon or phonon)  interacts with the electrons
within the dots. We set \begin{math} \hbar = 1 \end{math} in the following.

\subsection{The Hamiltonian}
An artificial `open' two-level system can be realized by
two quantum dots  (which for simplicity we call `left' and `right') coupled to each other by a tunnel barrier
and to a `left' source and `right' drain electron reservoir with chemical potentials $\mu_L$ and $\mu_R$.
In the following, we are only interested in the regime of high bias voltage $V=\mu_L-\mu_R>0$
We assume that the charging energy $U$
required to add an additional electron to the double dot
is much greater than $V$ (strong Coulomb blockade regime). Therefore,
electrons tunnel through the structure only from the left to the right.
Only one additional electron can tunnel into the double dot at a time, and the effective
Hilbert space of the electronic system can be defined by three states \cite{SN96,BK99}
`empty', `left', and `right', $|0\rangle = |N_{L},N_{R}\rangle$, $|L\rangle = |N_{L} +
1,N_{R}\rangle$, $|R\rangle = |N_{L},N_{R} + 1\rangle$.
The energies of the two states with an additional electron in the left (right) dot are
denoted as $\varepsilon_L$ ($\varepsilon_R$).
Higher excited states for both dots are assumed to lie outside the energy window 
\begin{eqnarray}
  \label{window}
\mu_L \gg \varepsilon_L,\varepsilon_R \gg \mu_R,
\end{eqnarray}
which defines the regime where nonlinear transport does not significantly depend 
on $\mu_L$ and $\mu_R$ at low temperatures. 
This corresponds to the standard situation \cite{SN96,Fujetal98,BK99} where
only the two lowest hybridized electronic states in the
double dot contribute to transport.

The Hamiltonian $H_D$ of the double dot is given in terms of the dot-operators
$n_L := |L\rangle
\langle L|$,
$n_R := |R\rangle \langle R|$,
and a term $H_T$ describing the tunneling between the left
and the right dot,
\begin{eqnarray}
H_D&=& H_A + H_T \equiv\varepsilon_L n_L + \varepsilon_R n_R +
T_c(P + P^\dagger), \nonumber\\
\quad P &=& |L\rangle \langle R|, \quad
P^\dagger = |R \rangle \langle L|.
\end{eqnarray}
The tunnel matrix element \begin{math} T_c \end{math} is
used to describe the strength of the tunneling process.

The Hilbert space of the bosonic system is spanned by the usual
number or Fock eigenstates $|n\rangle, n = 0,1,2,3,...$  
of the harmonic oscillator 
\begin{eqnarray}
H_p =\omega a^\dagger a   
\end{eqnarray}
with
frequency $\omega$ and creation/annihilation operators $a$ and $a^{\dagger}$ fulfilling $[a,a^\dagger]=1$.
The coupling between the electronic and the bosonic system is
described by four microscopic coupling constants,
\begin{equation}\label{ep_coupling}
 H_{ep} = (\alpha n_L + \beta n_R + \gamma^* P + \gamma P^\dagger)(a
 + a^\dagger).
\end{equation}
Here, we assume the coupling constants $\alpha,\beta$, and $\gamma$
as given parameters. Their precise form 
can be calculated from microscopic details such as
the many-body wave functions of the dot electrons. 
It should be noted that the interaction between the bosonic system and
the reservoirs is not considered here.

The two electron reservoirs are described by
\begin{equation}\label{The Electron Reservoirs}
H_{res} = \sum_k \varepsilon_k^L c_k^\dagger c_k + \sum_k
\varepsilon_k^R d_k^\dagger d_k,
\end{equation}
where the
sum is  over all
wave vectors $k$ in both
the left (L) and right (R) reservoirs and spin polarization
is assumed for simplicity .
The coupling between dot and reservoirs
is described with two dot operators
\begin{eqnarray}
s_L = |0\rangle \langle L|,\quad s_R = |0\rangle \langle R|
\end{eqnarray}
and the tunnel Hamiltonian
\begin{equation}\label{Reservoir-Dot Coupling}
H_V = \sum_k (V^L_k c_k^\dagger s_L + V^R_k d_k^\dagger s_R  + c.c)
\end{equation}
with tunnel matrix elements $V^L_k$ and $V_k^R$.

The total Hamiltonian is written as
\begin{equation}
H = H_0 + H_T + H_V + H_{ep},
\end{equation}
where
$H_0 = H_A + H_p + H_{res}$
is the free Hamiltonian
in the interaction picture introduced below.

\subsection{Equations of Motion}
The density matrix of the total system (dot, boson and reservoirs) is given by
the Liouville-von-Neumann equation
and formulated in the interaction picture with respect to $H_0$.
For the non-linear transport window Eq. (\ref{window}) considered here,
the chemical potentials in the equilibrium reservoirs
are such that the Fermi distributions in the left and the right lead  can well be
approximated by $f_L(\varepsilon)=1$ and $f_R(\varepsilon)=0$.
We treat the reservoir coupling term, $H_V$,
in second order perturbation theory, neglecting Kondo physics \cite{Hartmann_Kondo} throughout so that
transport can be described by a master equation \cite{SN96,GP96,Gur98,BK99}.
Transforming back into the Schr\"odinger picture
produces the following master equation for the reduced density operator of the
system (dot + boson),
\begin{eqnarray}\label{master1}
\frac{d}{dt}\rho(t) &=& -i[H_A + H_p + H_T + H_{ep},\rho(t)]\\
&-&\frac{\Gamma_L}{2}(s_L s_L^\dagger \rho(t) - 2s_L^\dagger
\rho(t)s_L + \rho(t)s_L s_L^\dagger)\nonumber\\
&-&\frac{\Gamma_R}{2}(s_R^\dagger s_R \rho(t) - 2s_R
\rho(t)s_R^\dagger + \rho(t)s_R^\dagger s_R)\nonumber\\
&-&\frac{\gamma_b}{2}(2a \rho a^\dagger - a^\dagger a \rho - \rho
a^\dagger a)\nonumber
\end{eqnarray}
The perturbation from the reservoir is given by the tunnel rates
\begin{eqnarray}\label{ratedef}
\Gamma_{L/R}= 2\pi \sum_k |V^{L/R}_{k}|^{2} \delta (\varepsilon_{L/R} -\varepsilon_{k}^{L/R}).
\end{eqnarray}
In the last line in Eq. (\ref{master1}), we introduced  a term
that describes damping of the  bosonic system at a rate
$\gamma_b$ \cite{Walls} corresponding to photon or phonon cavity losses.

Taking matrix elements of the master equation,
Eq. (\ref{master1}),
one obtains a system of linear equations for the matrix
elements of the density operator which are given explicitly in
appendix \ref{masterappendix}. In the stationary state of these equations, the time
derivatives are zero, and a coupled set of linear equations is
found.  In order to numerically solve these equations, the bosonic
Hilbert space has to be truncated at a finite number $N$ of boson
states. The total number of equations then is
$5 N^2 + 10  N + 5$,
remembering that there
is always an equation for $n=m=0$. The
numerical solution becomes
a standard matrix inversion, though
book-keeping of the matrix elements has to be done carefully.
The trace 
over electron and boson degrees of freedom has to
to obey the normalization condition
\begin{eqnarray}
\sum_n [ \langle n,0|\rho_{stat}|n,0 \rangle &+& \langle
n,R|\rho_{stat}|n,R \rangle\\
 &+& \langle n,L|\rho_{stat}|n,L\rangle] = 1\nonumber.
\end{eqnarray}
We mention that without this condition, the matrix becomes
singular, and cannot be solved.

\begin{figure}[t]
\includegraphics[width=0.5\textwidth]{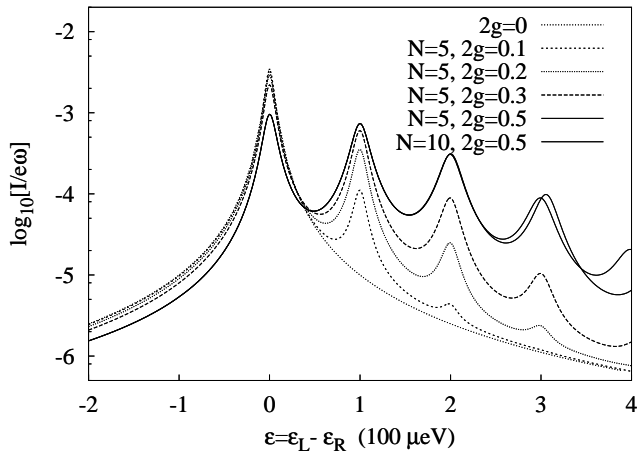}
\caption{Stationary current through double quantum dot coupled to
single boson mode with $\Gamma_L=\Gamma_R=0.1$, $T_c=0.01$,
$\gamma_b=0.05$ and varying $2g=\alpha=-\beta$. (electron-boson
coupling, cf. Eq. (\ref{ep_coupling})). $N$: number of boson states.} \label{fig1.eps}
\end{figure}

\section{Transport Properties}\label{section_transport}

The first quantity of interest that one can obtain from the matrix
elements is the stationary electron current through the double dot.

\subsection{Stationary Current: Numerical Results}
In the stationary state, the current operator is defined via
the rate of electrons tunneling from the left dot to the right dot,
\begin{equation}\label{currentdef}
\hat{I}=\frac{\partial}{\partial t} n_L= iT_C(P^\dagger - P) +
i(\gamma P^\dagger - \gamma^*P)(a + a^\dagger).
\end{equation}
Here and in the following we set the electron charge $-e=1$.
For a boson mode with wave vector ${\bf Q}$, the electron-boson coupling constants
can be expressed in terms of matrix elements of the left and right dot states,
$\alpha=\lambda \langle L| e^{i{\bf Qr}} |L\rangle$,
$\beta =\lambda \langle R| e^{i{\bf Qr}} |R\rangle$,
$\gamma = \lambda \langle L| e^{i{\bf Qr}} |R\rangle$,
where $\lambda$ is the microscopic constant for the interaction of the boson with the 
electron in 2D. 

\begin{figure}[t]
\includegraphics[width=0.5\textwidth]{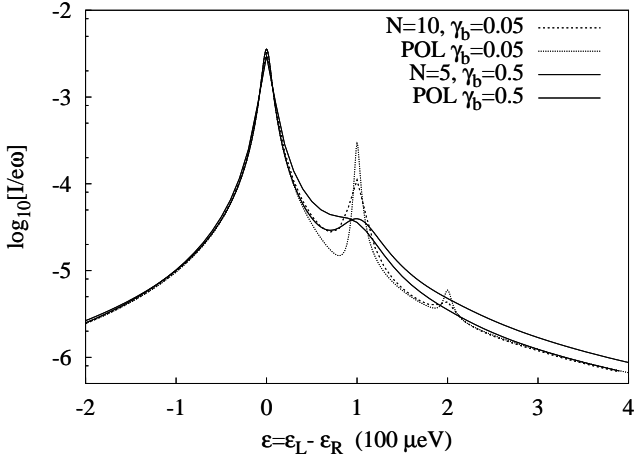}
\caption{Current as in Fig. \ref{fig1.eps} for coupling $2g=0.1$
and different boson damping $\gamma_b$, alongside corresponding
polaron transformation result, Eq. \ref{currentstat}.}
                \label{fig2.eps}
\end{figure}

Our formalism works for arbitrary choices of 
coupling parameters $\alpha$, $\beta$, $\gamma$, but for simplicity
we restrict ourselves to simple, non-trivial cases for the
numerical calculations. We set $\gamma=0$ which corresponds 
to neglecting non-diagonal terms that for 
relatively sharply peaked  electron densities in the dots only weakly contribute \cite{BV01}.
For a sharp electron density profile, one furthermore has
$\beta=\alpha e^{i{\bf Qd}}$ where ${\bf d}$ is the vector connecting
the left and right dot centers. Identical energy shifts  in both dots, corresponding to $\beta=\alpha$,
have no effect on transport. In the following, we choose ${\bf Qd}=\pi$ and therefore
electron-boson coupling constants
\begin{eqnarray}
  \alpha=-\beta \equiv 2g,\quad \gamma=0.
\end{eqnarray}
Furthermore, the tunneling rates between the reservoirs and
the dots are generally kept smaller than the energies of the dots
and boson states.  This produces clear and sharply defined
transport characteristics. We use the boson frequency as an energy scale and set $\omega=1$ in the following.
Using the typical phonon cavity energy $\hbar\omega \approx 100$ $\mu$eV 
from a recent experiment with a freestanding phonon cavity \cite{Hoeetal02} yields a typical 
electron current scale of $e\omega=24$ nA in Figs. \ref{fig1.eps} and \ref{fig2.eps}.

Fig. \ref{fig1.eps} illustrates the stationary current
as a function of the energy difference $\varepsilon$.
For zero boson coupling $g=0$, we reproduce the
Lorentzian resonant tunneling profile
\begin{equation}
I_{stat}=T_c^2\Gamma_R/[T_c^2(2+\Gamma_R/\Gamma_L) + \Gamma_R^2/4
+ \varepsilon^2]
\end{equation}
as first derived by Stoof and Nazarov~\cite{SN96}.

For finite electron-boson coupling
and positive $\varepsilon$, resonances appear  when
\begin{eqnarray}\label{rescondition}
  \varepsilon\equiv \varepsilon_L-\varepsilon_R = n \omega,
\end{eqnarray}
i.e., the energy gap of the dots becomes equal to
multiple integers of the boson energy. At these resonances,
electrons tunneling from the left
to the right dot can excite the boson system.
Note that for larger $T_c$, $\varepsilon=\varepsilon_L-\varepsilon_R$ in 
Eq. (\ref{rescondition}) has to be replaced by the energy difference
$\Delta\equiv \sqrt{\varepsilon^2+4T_c^2}$ between bonding and
anti-bonding state in the double dot.

The current profiles with differing damping rates,
Fig.(\ref{fig2.eps}),  illustrate that the system is extremely
sensitive even to very small `damping'. For small but finite boson
damping, we found that two profiles for $N=5$ and
$N=10$ match extremely closely. This illustrates that a finite
boson damping removes numerical problems due to the truncation of
the boson Hilbert space at a finite $N$. As expected, there are no
peaks at all on the absorption side of the profile
($\varepsilon<0$): the damped boson relaxes  to its ground state.

\subsection{Perturbation Theory in $T_c$: Polaron Transformation Method}\label{polaronsection}
An analytical expression for the stationary current can be obtained
from a polaron transformation and a polaron-transformed master equation
that is valid for small coupling $T_c$ between the two dots
\cite{BK99}. After tracing out the bosonic degrees of freedom one obtains
an equation of motion for the time-dependent expectation values of the operators
$n_{L/R}$ and $P^{(\dagger)}$ which can be transformed into Laplace-space,
\begin{eqnarray}\label{zequ}
\langle n_L\rangle (z)&=&-i\frac{T_c}{z}\left\{\langle P\rangle(z)-\langle P^{\dagger}
\rangle(z)\right\}\nonumber\\&+&
\frac{\Gamma_L}{z}\left\{1/z-\langle n_L\rangle(z)-\langle n_R\rangle(z)\right\}\nonumber\\
\langle n_R\rangle(z)&=&\phantom{-}i\frac{T_c}{z}\left\{\langle P\rangle(z)-
\langle P^{\dagger}\rangle(z)\right\}
-\frac{\Gamma_R}{z}\langle n_R\rangle(z)\nonumber\\
\langle P\rangle(z)&=&-iT_c\left\{\langle n_L\rangle(z)C_{\varepsilon}(z)
-\langle n_R\rangle(z)C_{-\varepsilon}^*(z)\right\}\nonumber\\&-&
\frac{\Gamma_R}{2}\langle P\rangle(z)C_{\varepsilon}(z).
\end{eqnarray}
Here, the coupling to the boson system enters through the correlation function
\begin{eqnarray}
  C_{\varepsilon}(z) &=&  \int_0^{\infty} dt e^{-z t+i\varepsilon t} C(t)\nonumber\\
  C(t) &=& \langle  X(t) X^{\dagger}(t=0) \rangle,\quad X = D\left( \frac{4g}{\omega}\right)
\end{eqnarray}
of the displacement operator of the  bosonic mode,
\begin{eqnarray}\label{displacedef}
D(\xi) = \exp (\xi a^{\dagger} - \xi^* a).
\end{eqnarray}
The equations (\ref{zequ}) can be solved algebraically. One then obtains the expectation value 
$ \langle I \rangle_{t\to\infty}$ of the stationary current from the
$1/z$-coefficient of the $I(z)$-expansion into a Laurent series \cite{Doetsch} for $z\to 0$,
\begin{eqnarray}\label{currentstat}
  & &\langle I \rangle_{t\to
  \infty}=T_c^2\frac{2\Re(C_{\varepsilon})+\Gamma_R|C_{\varepsilon}|^2}
  {|1+\frac{\Gamma_R}{2}C_{\varepsilon}|^2+2T_c^2B_{\varepsilon}}\\
B_{\varepsilon}&:=&
\Re\left\{(1+\frac{\Gamma_R}{2}C_{\varepsilon})\left[
 \frac{C_{-\varepsilon}}{\Gamma_R}+\frac{C_{\varepsilon}^*}{\Gamma_L}\left(1+\frac{\Gamma_L}{\Gamma_R}\right)\right]\right\}\nonumber,
\end{eqnarray}
where $C_\varepsilon\equiv C_\varepsilon(z\to 0)$.

The correlation function, $C(t)$, enters via 
a factorization assumption in the polaron-transformed master equation and
has to be calculated from an effective density operator $\rho_{\rm B}$ of the bosonic mode.
For dissipative pseudospin-boson systems with a continuous spectrum of 
modes and a thermal equilibrium for $\rho_{\rm B}$, reasonable results 
can be obtained within this method for small coupling, in particular in comparison with
perturbation theory \cite{BV02,BV01}. However, in the present case with just one single mode
it is not clear whether or not  a factorization into an effective dot and an effective boson density
operator yields reasonable results, 
and in principle one should at least determine $\rho_{\rm B}$ self-consistently.
For small couplings $g$, however, the assumption of a decoupled time-evolution of the boson should work well,
and we proceed with a master equation for $\rho_{\rm B}$,
\begin{eqnarray}
  \frac{d}{dt}\rho_{\rm B}(t) &=& -i[\omega a^{\dagger}a,\rho_{\rm B}] \nonumber\\
&-& \frac{\gamma_b}{2}\left( 2a\rho_{\rm B} a^{\dagger} -
a^{\dagger} a \rho_{\rm B} - \rho_{\rm B} a^{\dagger} a \right)
\end{eqnarray}
that describes the free time-evolution of the boson in presence of boson
damping with damping constant $\gamma_b>0$. Assuming the boson initial condition to be
its ground state, one obtains
\begin{eqnarray}
  C(t) = \exp\left\{-|\xi|^2\left( 1- e^{-\left(\frac{\gamma_b}{2}+i\omega\right) t}\right)\right\},
\quad \xi=\frac{4g}{\omega}.
\end{eqnarray}
Fig. (\ref{fig2.eps}) shows that the analytical expression for the
current,  Eq.(\ref{currentstat}), compares quite well with the 
numerical results. In particular, this demonstrates that the
formulation in terms of a general boson correlation function
$C(t)$ works reasonably well not only for boson systems with
continous mode spectrum \cite{BK99,BV01} but  also for the single
mode case discussed here, at least for small coupling constants $g$. 
We point out that the existence of a finite boson damping is crucial here and that
for larger $g$, the comparison becomes worse, as expected.

\section{Boson Distribution}\label{section_boson}
Turning now to properties of the bosonic system, our
primary interest is how the boson mode can be controlled
via parameters of the electron sub-system such as the bias $\varepsilon$
and the tunnel couplings $\Gamma_{R/L}$.
In the stationary state, the reduced density operator of the boson is
\begin{eqnarray}\label{rhobdef}
  \rho_b \equiv \lim_{t\to \infty} {\rm Tr_{\rm dot}}\rho(t)=
\lim_{t\to \infty}\sum_{i=0,L,R}\rho^{ii}(t).
\end{eqnarray}
\subsection{Boson state detection}
Before discussing the boson state $\rho_b$ we describe a possible
experimental scheme to directly detect its properties. The basic idea is to use
another nearby double quantum dot as detector of the boson, similar to the scheme for
detecting quantum noise in mesoscopic conductors as proposed by Aguado and Kouwenhoven \cite{AK00}.

The detector consists of a double quantum dot very similar to the one discussed above. 
The boson state gives a contribution to the inelastic, stationary  current 
\begin{eqnarray}\label{Id}
 I_d(E) \approx T_d^2 P(E) &=& 2T_d^2 \Re \int_0^{\infty} dt e^{iEt}C_d(t)\\
C_d(t) &=& {\rm Tr}\left[ \rho_b X_d(t)X_d^{\dagger}\right],
\quad X_d=D\left(\frac{4g_d}{\omega}\right),\nonumber
\end{eqnarray}
through the 
detector in lowest order in the detector  tunnel coupling $T_d$. Here, $E$ denotes the 
detector dot energy difference, 
$D$ again denotes the unitary displacement operator, Eq. (\ref{displacedef}), 
$g_d$ the coupling constant in the detector double dot, and $\rho_b$ the boson state. 
Eq. (\ref{Id}) describes the detector current $I(E)$ resulting from a given boson state
$\rho_b$ as calculated in the next section.

Before discussing our  numerical results for $\rho_b$,
it is illustrative to look at a few special cases.
For example, if the boson state $\rho_b$ describes a thermal equilibrium at temperature
$T$, one obtains the inelastic current spectrum with absorption $E<0$ and emission $E>0$ branches \cite{BK99}, 
as has been observed by Fujisawa et al. \cite{Fujetal98} for the case of equilibrium  (multimode) phonons.

If $\rho_b=|n\rangle \langle n|$ describes a pure, $n$-boson number state, the function
$C_d(t)$ in the detector current, Eq. (\ref{Id}), becomes ($\xi = 4g_d/\omega$)
\begin{eqnarray}\label{Cdt1}
  C_d(t)&=& \langle n | D(\xi e^{i\omega t}) D (-\xi) |n\rangle =\\
&=& e^{-i |\xi|^2 \sin \omega t} e^{-\frac{1}{2}\left|\xi\left(1-e^{i\omega t}\right)\right|^2 }
L_n^0\left(\left|\xi(1-e^{i\omega t})\right|^2\right),\nonumber
\end{eqnarray}
where $L_n^0$ is a Laguerre polynomial, cf. appendix \ref{appendixuseful}. 
Here we assume that the boson time-evolution 
is undamped in the detector, i.e.
governed by the boson Hamiltonian $H_b=\omega a^{\dagger} a$. 
Expanding up to lowest order in $\xi=4g_d/\omega$ and using $L_n^0(x)=1-nx + O(x^2)$, one obtains
from Eq.(\ref{Id}) and Eq.(\ref{Cdt1})
\begin{eqnarray}
  I_d(E) &\approx& T_d^2 2\pi \left[\left\{1-|\xi|^2(2n+1)\right\}\delta(E)\right.\\
 &+&\left. |\xi|^2\left\{n\delta(E+\omega)+(n+1)\delta(E-\omega)\right\}\right]\nonumber
\end{eqnarray}
Clearly, the height of the inelastic current peaks is determined by the quantum number
$n$, i.e. the  absorption peak at $E=-\omega<0$ scales with $n$, and the (stimulated) emission
peak at $E=\omega>0$ scales with $n+1$. 

The above example can be even generalised. In appendix 
\ref{bosonappendix} we show how to reconstruct an arbitrary boson state
$\rho_b=\sum_{nm}\rho_{nm}|n\rangle \langle m|$ from the function $P(E)$
(or equivalently the stationary detector current spectrum $I_d(E)$, Eq. (\ref{Id})). 
This demonstrates that the properties of the boson state discussed in the following can
be directly related to an experimentally accessible quantity.


\subsection{Numerical results}
We first analyse the boson mode via moments of its distribution function
such as occupation number and  Fano factor, before
discussing the full distribution function in the Wigner function representation.

\begin{figure}[t]
\includegraphics[width=0.5\textwidth]{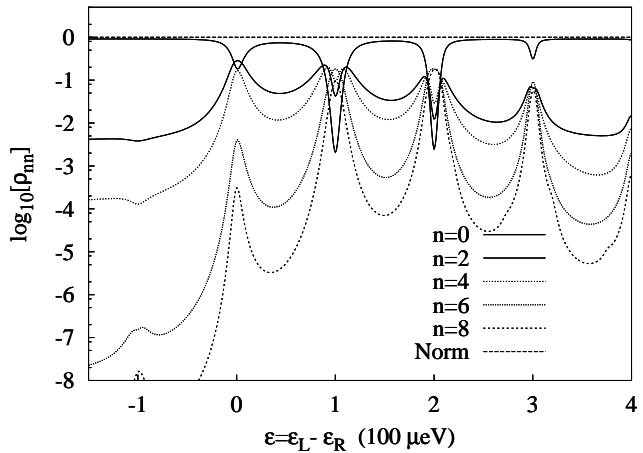}
\caption{Boson profiles n=0,2,4...10 for $\gamma_b=0.0001$,
$N=10$, $2g=\alpha=-\beta=0.3$, $10\times
T_C=\Gamma_L=\Gamma_R=0.1$.} \label{smalldampingboson}
\end{figure}
The boson occupation probability $p_n=(\rho_b)_{nn}$ is the probability for the
boson to be in one particular number state $n$. 
Furthermore, the boson Fano factor $F^Q$ is calculated from the boson occupancy
as the variance of boson number over the
mean number of bosons,
\begin{equation}
F^Q=\frac{\langle \hat{n}^2 \rangle - \langle \hat{n}
\rangle^2}{\langle \hat{n} \rangle}.
\end{equation}
Similarly, the variance
$\Delta x^2 = \langle x^2\rangle - \langle x \rangle^2$ and
$\Delta p^2 = \langle p^2\rangle - \langle p \rangle^2$
of the position and a momentum coordinate
\begin{eqnarray}\label{xpdef}
  x=\frac{(a+a^\dagger)}{\sqrt{2}},\quad p=\frac{i(-a +
a^\dagger)}{\sqrt{2}}
\end{eqnarray}
can be obtained from $\rho_b$,
cf. Eq. (\ref{variancesdef}) in appendix \ref{masterappendix}. They provide an
indication of the quantum fluctuations in the boson states, and prove useful in
comparison with the Fano factor profile. It should also be
apparent that $\Delta x^2 \Delta p^2 \geq \frac{1}{4}$.

We have checked that the probability $p_n$ fulfills the
normalisation $\sum_n p_n=1$. For accurate results the occupation
probability should tend to 0 as $n \rightarrow N$, where $N$ is
the dimension of the truncated boson Hilbert space.  This
condition is easily achieved in presence of a finite boson damping
rate, $\gamma_b>0$. In this case, it is  possible to obtain a
boson distribution which is centered around the lower boson number
states, and is excited when a resonant interaction with the
electron occurs.

It is more illustrative to first discuss a weakly damped case, as
shown in Fig. \ref{smalldampingboson}.  We can see quite clearly
how at the positive resonance energies, the occupation
probabilities of the states spreads into the higher number states.

\begin{figure}[t]
\includegraphics[width=0.5\textwidth]{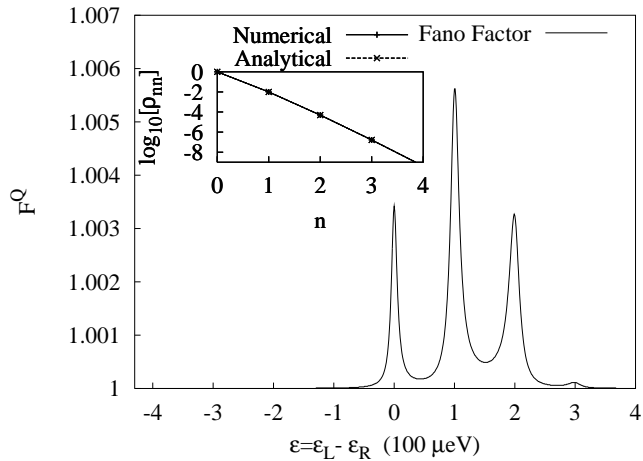}
\caption{Fano versus $\varepsilon$ profile for
$2g=\alpha=-\beta=0.1$, $N=10$,   $\gamma_b=0.05$, $10\times
T_C=\Gamma_L=\Gamma_R=0.1$.  Inset: Comparison between
coherent  state, Eq.(\ref{coherentanalytical}), 
and numerical result for boson distribution $p_n$ at $\varepsilon=-4$.}
\label{damping_fano}
\end{figure}

The Fano factor $F^Q$ illustrates
fluctuation properties of the boson mode
with $F^Q=1$ corresponding to a Poissonian boson number distribution and
$F^Q<1$ ($F^Q>1$) to a sub (super)-Poissonian distribution.
Fig. \ref{damping_fano} illustrates the Fano factor vs. $\varepsilon$ for
the strongly damped case.

This profile suggests the 
reduced boson states are coherent (Poissonian) for $\varepsilon<0$, and
super-Poissonian at the resonance energies, $\varepsilon =
\omega,2\omega, ...$.

\subsection{Coherent state and $x$-$p$ variances}
For $\varepsilon \ll 0$, nearly no current is flowing, the
electron is predominatly localized in the left dot, and one can
approximate the operator $\sigma_z=| L\rangle \langle L| -
|R\rangle \langle R|$ by its expectation value $\langle \sigma_z
\rangle =1$. Then, the boson system is effectively described by
\begin{eqnarray}
  H_{\rm eff} = 2g(a+a^{\dagger}) + \omega a^{\dagger}a
\end{eqnarray}
which is a shifted oscillator, the ground state $|GS\rangle$ of
which is the coherent state $|GS\rangle
=|-2g/\omega\rangle$ with $a|z\rangle = z
|z\rangle$. This can be seen by introducing new operators
$b:=a+2g/\omega$ whence $H_{\rm eff} =  \omega b^{\dagger}b-4g^2/\omega$
and $b|GS\rangle = 0$. In this regime, we have 
\begin{eqnarray}\label{coherentanalytical}
 \rho_b  \approx |z\rangle \langle z|,\quad z=-2g/\omega,
\end{eqnarray}
and $p_n=|\langle n|GS\rangle|^2$ is given
by a Poisson distribution, $p_n =|z|^{2n}e^{-|z|^2}/n!$.
One can plot this against the numerically obtained $p_n$ and check that
the distribution is indeed Poissonian for small $\varepsilon \ll 0$.  This is
represented in the inset of Fig. \ref{damping_fano} in detail for
the first 4 boson Fock (number) states.

\begin{figure}[t]
\includegraphics[width=0.5\textwidth]{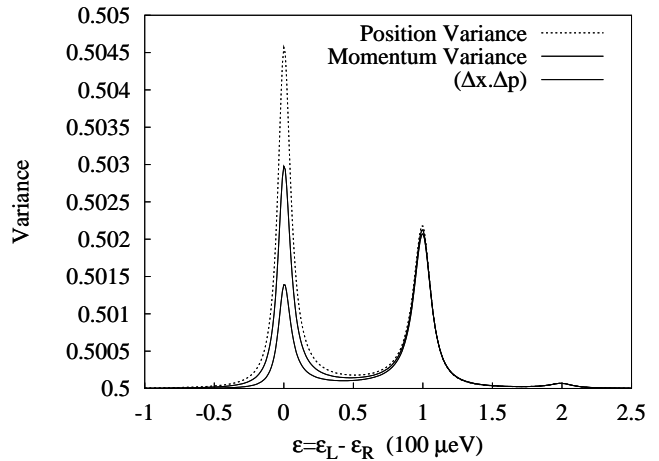}
\caption{Position variance, momentum variance, and an estimation
of the combined uncertainty. $2g=\alpha=-\beta=0.1$, $N=10$,
$\gamma_b=0.05$,  $10\times T_C=\Gamma_L=\Gamma_R=0.1$.}
\label{variance}
\end{figure}

The quantum fluctuations of the boson system
is described by the variance of the position and momentum
operators.
Fig. \ref{variance} presents  a direct comparison of the position and
momentum variance of a strongly damped system. The
uncertainty principle holds,  $\Delta x \Delta p \geq
\frac{1}{2}$, and thus $\Delta x^2 \Delta p^2 \geq \frac{1}{4}$.
The state with minimum uncertainty occurs as expected, for
$\varepsilon_L < \varepsilon_R$, with $\Delta x^2 \Delta p^2 =
\frac{1}{4}$.  This is a reassuring result, as it suggests both
that the boson is coherent in this area, and that the Fano
factor's Poissonian distribution for this coherent state is
correct. 
We mention that since both $\Delta x>0.5$ and $\Delta p >0.5$, we find no squeezing of the boson
mode for the parameters checked here.

\subsection{Wigner Function}
The Wigner function of the density operator ${\rho_b}$ for a bosonic
mode $a^{\dagger}$ is a representation of $\rho_b$ in $x$-$p$-space, cf. Eq.(\ref{xpdef}). 
It is defined as \cite{CG69}
\begin{eqnarray}
  W(x,p):=\frac{1}{\pi}{\rm Tr}\left(\rho_b D(2\alpha) U_0 \right),  \quad \alpha=\frac{x+ip}{\sqrt{2}},
\end{eqnarray}
where $D(\alpha):=\exp[\alpha a^{\dagger} - \alpha^* a]$, Eq.(\ref{displacedef}), is a
unitary displacement operator and $U_0:=\exp[i\pi a^{\dagger} a]$
parity operator for the boson \cite{CV98}. $W(x,p)$ is a symmetric Gaussian for 
a pure coherent boson state and a symmetric Gaussian multiplied with a polynomial for a 
pure number state \cite{Walls}. In our model, the shape of $W(x,p)$ therefore indicates how 
close to these limiting cases the actual stationary state $\rho_b$ of the boson, Eq.(\ref{rhobdef}), is.
In particular, 
this is a convenient way to represent the `steering' of the boson mode 
when external parameters (like $\varepsilon$ or $\Gamma_{R/L}$) are changed.

Using the Fock state basis $\{|n\rangle,
n=0,1,2,... \}$ and $U_0=\sum_{n=0}^{\infty}(-1)^n |n\rangle
\langle n |$, we find
\begin{eqnarray}
  W(x,p)=\frac{1}{\pi}\sum_{n,m=0}^{\infty}(-1)^n \langle n |\rho_b|m \rangle \langle m| D(2\alpha)|n\rangle.
\end{eqnarray}
It is useful to split the sum into diagonal and non-diagonal parts,
to use $\langle m |\rho_b|n\rangle $ = $\langle n |\rho_b|m\rangle ^*$
and ($m\ge n$)
\begin{eqnarray}\label{Laguerre}
\langle m | D(\alpha) | n\rangle &=&
\sqrt{\frac{n!}{m!}}\alpha^{m-n}e^{-\frac{1}{2}|\alpha|^2}L_{n}^{m-n}\left(|\alpha|^2\right)\\
&=& \langle m| D^{\dagger}(\alpha)|n\rangle^*
=(-1)^{m-n} \langle m| D(\alpha)|n\rangle^*\nonumber,
\end{eqnarray}
where again $\alpha=({x+ip})/{\sqrt{2}}$ and $L_{n}^{m-n}$ is a Laguerre polynomial, cf. appendix
\ref{appendixuseful}.
This leads to
\begin{eqnarray}\label{Wxpresult}
\lefteqn{ W(x,p) = \frac{1}{\pi}\sum_{n=0}^{\infty}(-1)^n
\langle n |\rho_b|n\rangle \langle n| D(2\alpha)|n\rangle }\\
&+&\frac{1}{\pi}\sum_{n=0}^{\infty} \sum_{m=n+1}^{\infty}(-1)^n
2{\rm Re} \left[\langle n |\rho_b|m \rangle \langle m|
D(2\alpha)|n\rangle\right].\nonumber
\end{eqnarray}

\begin{figure}[t]
\psfrag{eps}{$\varepsilon$}
\psfrag{=}{$=$}
\psfrag{0.00}{ $0.00$}
\psfrag{0.25}{ $0.25$}
\psfrag{0.50}{ $0.50$}
\psfrag{0.75}{ $0.75$}
\psfrag{1.00}{ $1.00$}
\psfrag{1.25}{ $1.25$}
\psfrag{1.50}{ $1.50$}
\psfrag{1.75}{ $1.75$}
\psfrag{2.00}{ $2.00$}
\psfrag{2.25}{ $2.25$}
\includegraphics[width=0.5\textwidth]{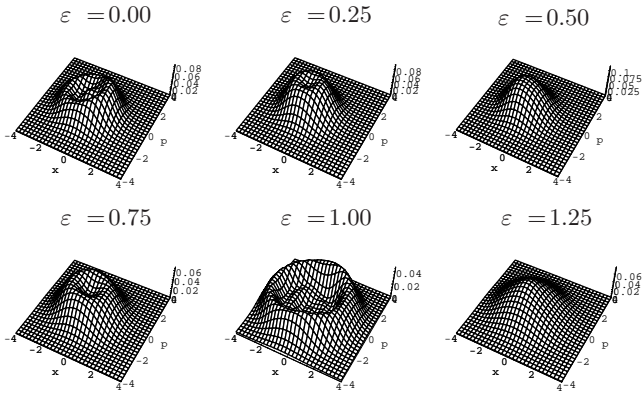}
\caption[]{\label{fig6.eps}Wigner distribution functions for the
bosonic mode. Parameters are $\Gamma_L=\Gamma_R=T_c=0.1$,
$\gamma_b =0.005$, $g=0.2$, $N=20$. Stationary current resonances occur at
$\varepsilon=0.0, 1.0, 2.0,...$.}
\end{figure}
Using our numerical results for the stationary state $\rho$ of the coupled system, 
we easily obtain  Wigner functions from Eq.(\ref{rhobdef}) and Eq.(\ref{Wxpresult}) for different values of 
the energy difference $\varepsilon$ between the two dots as shown in Fig. (\ref{fig6.eps}).
Between two resonance energies $\varepsilon=n\omega$, $W(x,p)$ closely resembles a Gaussian. At and close
to the resonance energies, the distribution spreads out in rings around the origin, which is 
consistent with the increased Fock state occupation numbers 
and the increased position and momentum variances at these energies.

\section{Discussion}\label{section_discussion}
Our investigation is based  on a
numerical method which allows the density matrix of the boson-dot system to be solved for
arbitrary dot parameters ($\varepsilon$, $T_c$, $\Gamma_{L/R}$) and boson damping rate $\gamma_b$.
Since we have to truncate the boson Hilbert space, the electron-boson coupling constant $g$ 
has to be restricted to small values $g \lesssim 1$. In contrast to 
the determination of the (pure) eigenstates of an isolated dot-boson system
(Rabi-Hamiltonian), the numerical effort for our mixed state (density operator) is much bigger here.
Although
not discussed in this paper, we suggest that 
the strong coupling regime could be reached numerically by a polaron transformation
of the master equation \cite{BK99} without the factorization assumption 
employed in section \ref{polaronsection}.
Fortunately, at present the small coupling regime seems to be valid for 
experimental situations with quantum dots \cite{Fujetal98,Taretal99,BK99,Qinetal01}.

Our results suggest that there is no resonant
interaction on the $\varepsilon<0$ side of the current
peak as long as the boson system is damped and any
many-body excited electron states can be ignored.
For $\varepsilon>0$  we have found strong excitations of the boson
mode occuring at resonances given by multiples $n\omega$ of the boson
frequency $\omega$. These correspond to the photo-satellite peaks in resonant
tunneling in electromagnetic AC fields \cite{Oosetal98,SN96} which in the classical case, however,
appear on both sides of the main resonance $n=0$ in contrast to the quantum case considered here.
The resonances at $\varepsilon=n\omega$  can thus be interpreted as the emission of phonons (photons) 
by the electron as it tunnels through the dot. It should be possible to detect them in the
stationary current through double quantum dots in `phonon resonators'.
With increasing electron-boson coupling, the visibility of these
side peaks increases whereas the main resonance at $\varepsilon = 0$ is reduced.

We have discussed that an additional double dot 
can serve as a detector of the stationary boson state $\rho_b$, if 
its energy difference $E$  is tuned independently. 
In the single mode case considered here, 
the stationary detector current spectrum $I_d(E)$ is uniquely related to the components of
$\rho_b$ in the number state basis.
In order to characterize the 
stationary boson state itself, we have investigated the Fano factor and the quantum
fluctuations in the quadrature amplitudes of the boson. For $\varepsilon \ll 0$ the
boson state is perfectly described as a coherent state whereas
for $\varepsilon \gtrsim 0$ the interaction with the quantum dot produces
occupied excited states in the boson mode that could best be visualized by
a broadening of the phase space (Wigner) distribution function of the  boson.


Although in this paper we have only calculated stationary properties, our method can be extended 
to calculate the noise-spectrum via the quantum regression theorem within the master equation
framework as well \cite{Walls}. It remains a task for the future to 
analyze the relation between the boson quantum fluctuations and the current noise in detail.

\begin{acknowledgments}
We acknowledge inspiring discussions with  A. Vourdas (who suggested
the analysis of the $x$ and $p$ variances) and C. Emary.
This work was supported by projects EPSRC GR44690/01, DFG Br1528/4-1,
the WE Heraeus foundation and the UK Quantum Circuits Network.
\end{acknowledgments}

\begin{appendix}
\section{Matrix elements of the density operator}\label{masterappendix}
The matrix elements of the dot-boson density operator $\rho$ are defined as
\begin{eqnarray}
\rho_{nm}^i &:=& \langle n,i|\rho|i,m \rangle,\quad i=0,L,R \nonumber\\
\rho_{nm}^{ij} &:=& \langle n,i|\rho|j,m \rangle,\quad i,j=0,L,R,
\end{eqnarray}
where $n$ and $m$ refer to boson Fock states.

The matrix elements for the empty dot state obey
\begin{eqnarray}
\frac{d}{dt}\rho_{nm}^0 &=& [i\omega(m - n)\rho_{nm}^0]
-\Gamma_L[\rho_{nm}^0] + \Gamma_R[\rho_{nm}^R]\\
&+&\frac{\gamma_b}{2}[2 \rho_{n+1,m+1}^0\sqrt{n+1}\sqrt{m+1} -
\rho_{nm}^0(n+m)],\nonumber
\end{eqnarray}
those for the right dot state
\begin{eqnarray}
\frac{d}{dt}\rho_{nm}^R &=& i[\omega(m - n)\rho_{nm}^R + T_c(\rho_{nm}^{RL} - \rho_{nm}^{LR})\\
&+&\beta(\sqrt{m}\rho_{n m-1}^R + \sqrt{m+1}\rho_{n
m+1}^R\nonumber\\
&-&\sqrt{n}\rho_{n-1 m}^R - \sqrt{n+1}\rho_{n+1 m}^R)\nonumber\\
&+&\gamma^*(\sqrt{m}\rho_{n m-1}^{RL} + \sqrt{m+1}\rho_{n
m+1}^{RL})\nonumber\\
&-& \gamma(\sqrt{n}\rho_{n-1 m}^{LR} + \sqrt{n+1}\rho_{n+1
m}^{LR})] - \Gamma_R[\rho_{nm}^{R}]\nonumber\\
&+& \frac{\gamma_b}{2}[2 \rho_{n+1,m+1}^R\sqrt{n+1}\sqrt{m+1} -
\rho_{nm}^R(n+m)],\nonumber
\end{eqnarray}
and for the left dot state
\begin{eqnarray}
\frac{d}{dt}\rho_{nm}^L &=& i[\omega(m - n)\rho_{nm}^L + T_c(-\rho_{nm}^{RL} + \rho_{nm}^{LR})\\
&+&\alpha(-\sqrt{n+1}\rho_{n+1 m}^L  - \sqrt{n} \rho_{n-1 m}^L\nonumber\\
&+&\sqrt{m}\rho_{n m-1}^L + \sqrt{m+1}\rho_{n m+1}^L)\nonumber\\
&+& \gamma^*(-\sqrt{n+1}\rho_{n+1 m}^{R} - \sqrt{n}\rho_{n-1
m}^{RL})\nonumber\\
&+&\gamma(\sqrt{m}\rho_{n m-1}^{LR}+\sqrt{m+1}\rho_{n m+1}^{LR} )]
+ \Gamma_L[\rho_{nm}^{0}]\nonumber\\
&+&\frac{\gamma_b}{2}[2 \rho_{n+1,m+1}^L\sqrt{n+1}\sqrt{m+1} -
\rho_{nm}^L(n+m)].\nonumber
\end{eqnarray}
The equation of motion for the  off-diagonal elements 
$\rho_{nm}^{RL}$ is
\begin{eqnarray}
\frac{d}{dt}\rho_{nm}^{RL} &=& i[\rho_{nm}^{RL}[\omega(m - n)
+ (\varepsilon_L - \varepsilon_R)] \\
&+&\beta(-\sqrt{n+1}\rho_{n+1 m}^{RL} - \sqrt{n}\rho_{n-1 m}^{RL})\nonumber\\
&+&\alpha(\sqrt{m+1}\rho_{n m+1}^{RL} + \sqrt{m}\rho_{n m-1}^{RL})\nonumber\\
&+&\gamma(-\sqrt{n+1}\rho_{n+1 m}^{LL} - \sqrt{n}\rho_{n-1 m}^{LL}\nonumber\\
&+&\sqrt{m+1}\rho_{n m+1}^{RR} + \sqrt{m}\rho_{n m-1}^{RR})\nonumber\\
&+&T_c(\rho_{nm}^{RR} - \rho_{nm}^{LL})]
-\frac{\Gamma_R}{2}[\rho_{nm}^{RL}]\nonumber\\
&+&\frac{\gamma_b}{2}[2 \rho_{n+1,m+1}^{RL} \sqrt{n+1}\sqrt{m+1} -
\rho_{nm}^{RL}(n+m)]\nonumber.
\end{eqnarray}

Taking the trace over all the boson and electron states, the expression
Eq. (\ref{currentdef}) for the electron current operator  reads
\begin{eqnarray}
\langle \hat{I} \rangle &=& \sum_n iT_C [\rho_{n,n}^{LR} -
\rho_{n,n}^{RL}]\\
&+& \sum_n i[\gamma (\rho_{n,n-1}^{LR}\sqrt{n} +
\rho_{n,n+1}^{LR}\sqrt{n+1}) \nonumber\\
&-& \gamma^*( \rho_{n,n-1}^{RL}\sqrt{n} +
\rho_{n,n+1}^{RL}\sqrt{n+1})]\nonumber.
\end{eqnarray}
The variances of the boson position and  momentum coordinate, Eq.(\ref{xpdef}),
are obtained by performing the trace over the dot variables $0,L,R$,
\begin{eqnarray}\label{variancesdef}
\Delta x^2 &=&
\frac{1}{2}{\rm Tr}_{\rm dot}\sum_n(\sqrt{n}\sqrt{n-1}\rho_{n,n-2} \\
&+& (2n+1) \rho_{n,n} + \sqrt{n+1}\sqrt{n+2}\rho_{n,n+2}) \nonumber\\
&-&\frac{1}{2}[{\rm Tr}_{\rm dot}\sum_n[\sqrt{n}\rho_{n,n-1} +
\sqrt{n+1}\rho_{n,n+1}]]^2,\nonumber\\
\Delta p^2 &=& \frac{1}{2}{\rm Tr}_{\rm dot}\sum_n(-\sqrt{n}\sqrt{n-1}\rho_{n,n-2}\\
&+&(2n+1) \rho_{n,n} - \sqrt{n+1}\sqrt{n+2}\rho_{n,n+2})\nonumber\\
&-& \frac{1}{2}[{\rm Tr}_{\rm dot}\sum_n[-i\sqrt{n}\rho_{n,n-1} +
i\sqrt{n+1}\rho_{n,n+1}]]^2.\nonumber
\end{eqnarray}

\section{Displacement operator, coherent states}\label{appendixuseful}
Here, we summarize some useful properties of the unitary displacement operator
\begin{eqnarray}
D(z) &\equiv& e^{z a^{\dagger} - z^* a}= \left(D^{\dagger}(z)\right)^{-1}=D^{\dagger}(-z)\nonumber\\
&=&e^{-\frac{1}{2}|z|^2}
e^{z a^{\dagger}} e^{- z^* a}=e^{\frac{1}{2}|z|^2}e^{- z^* a}
e^{z a^{\dagger}},
\end{eqnarray}
where $z$ is a complex number and we used the
operator exponential $e^{A+B}=e^Ae^Be^{-(1/2)[A,B]}$ for $[[A,B],A]=[[A,B],B]=0$,
cf. also \cite{Walls}.
A coherent boson state $|z\rangle$ is defined as eigenstate of the 
annihilation operator, $a  |z\rangle = z |z\rangle$, where $z$ is a complex number.
It can be generated from the boson vacuum $|0\rangle$ as
\begin{eqnarray}
  |z\rangle = D(z) |0\rangle.
\end{eqnarray}
Extremely useful is the relation
\begin{eqnarray}
  D(\alpha+\beta) = D(\alpha) D(\beta) e^{-i\Im (\alpha\beta^*)}
\end{eqnarray}
for arbitrary complex numbers $\alpha$, $\beta$.
Coherent state matrix elements of $D(\alpha)$ follow as
\begin{eqnarray}\label{Drule}
  \langle \beta |D(\alpha)| \beta \rangle = e^{-\frac{1}{2}|\alpha|^2}
e^{2i\Im (\alpha \beta^*)}.
\end{eqnarray}
Number state matrix elements can be obtained using $|n\rangle \equiv (1/\sqrt{n!})(a^{\dagger})^n|0\rangle$,
\begin{eqnarray}
 & & \langle m | D(\alpha) | n\rangle
= e^{\frac{1}{2}|\alpha|^2}\times\\
&\times& 
\frac{1}{\sqrt{n!m!}}\frac{\partial^{n+m}}{\partial z_1^m \partial z_2^n}
\left.\langle 0|e^{(z_1-\alpha^*)a} e^{(z_2+\alpha)a^{\dagger}}|0\rangle
\right|_{z_1=z_2=0}.\nonumber
\end{eqnarray}
With $\langle 0|e^{(z_1-\alpha^*)a} e^{(z_2+\alpha)a^{\dagger}}|0\rangle$
$= e^{(z_1-\alpha^*)(z_2+\alpha)}$, for $m\ge n$ the differentiation yields
\begin{eqnarray}
   \langle m | D(\alpha) | n\rangle&=&e^{\frac{1}{2}|\alpha|^2} \frac{1}{\sqrt{n!m!}}\times\nonumber\\
&\times&  \left.\frac{\partial^{n}}{\partial z_2^n}(z_2+\alpha)^m e^{-(z_2+\alpha)\alpha^*}
\right|_{z_2+\alpha}.
\end{eqnarray}
Comparison with the generating function of the Laguerre polynomials \cite{Gradstein} yields Eq.(\ref{Laguerre}).
A corresponding expression can be derived for $n\ge m$.

\section{Relation between $P(E)$ and stationary boson state $\rho_b$}\label{bosonappendix}
Again we assume that the boson time-evolution is undamped in the detector, i.e.
governed by the  boson Hamiltonian $H_b=\omega a^{\dagger} a$. 
Expressing an arbitrary boson state $\rho_b$ in the number state basis,
$\rho_b=\sum_{nm}\rho_{nm}|n\rangle \langle m|$, Fourier transforming the function $P(E)$, Eq.(\ref{Id}),
yields ($\xi=4g_d/\omega$)
\begin{eqnarray}\label{P_t}
\tilde{P}(t)&\equiv & \frac{1}{2\pi}\int_{-\infty}^{\infty}dE e^{-iEt}P(E) = C_d(t)+C_d^*(-t)\nonumber\\
&=& \sum_{nm}\rho_{nm}c_{nm}(t) \\
c_{nm}(t)&\equiv& 
\langle   m| D(\xi e^{i\omega t}) D (-\xi) |n\rangle
+ \langle m| D (\xi) D(-\xi e^{-i\omega t}) |n\rangle.\nonumber
\end{eqnarray}
The functions $c_{nm}(t)$ can be calculated analytically,
using Eq. (\ref{Drule}) and the matrix elements $\langle m|D(z)|n\rangle$ of the unitary displacement operator, Eq. (\ref{Laguerre}).
They are periodic in $2\pi/\omega$, and from integrating Eq.(\ref{P_t}) 
over one period one obtains a linear relation between the
Fourier coefficients $\tilde{P}^k$ and $c_{nm}^k$,
\begin{eqnarray}\label{tensor}
  \tilde{P}^k &=& \sum_{nm}c_{nm}^k \rho_{nm}\\
\tilde{P}^k&\equiv& \frac{\omega}{2\pi}\int_0^{\frac{2\pi}{\omega}} \tilde{P}(t)e^{ik\omega t}dt\nonumber\\
c_{nm}^k&\equiv& \frac{\omega}{2\pi}\int_0^{\frac{2\pi}{\omega}}c_{nm}(t)e^{ik\omega t}dt.\nonumber
\end{eqnarray}
The $\tilde{P}^k$ have to be determined by numerical integration from 
the (experimentally given) $P(E)$. 
Regarding 
$(nm)$ as a single index, Eq.(\ref{tensor}) is a linear equation that can then be 
solved for the coefficients $\rho_{nm}$ of the boson state by inverting
the matrix $c_{nm}^k$, the coefficients of which are
given as Fourier coefficients of known expressions. In practical terms,
the number of boson states as well as the number of Fourier 
coefficients have to be restricted in order to make this
inversion feasable. 

Eq. (\ref{tensor}) establishes the
relation between $P(E)$ (or, via Eq. (\ref{Id}), the detector current $I_d=T_d^2P(E)$ in lowest 
order $T_d$) and an arbitrary single mode  boson state $\rho_b$.

\end{appendix}


\end{document}